\documentclass{Interspeech2024}

\usepackage{multirow}
\usepackage[table,xcdraw]{xcolor}
\usepackage{hyperref}
\usepackage{cite}
\usepackage{tablefootnote}

\interspeechcameraready

\newcommand{\newpar}{\vspace{1mm}\noindent\textbf}

\newcommand{\poli}{$^{\clubsuit}$}
\newcommand{\sfh}{$^{\diamondsuit}$}
\newcommand{\sgbh}{$^{\heartsuit}$}
\newcommand{\unito}{$^{\spadesuit}$}

\usepackage{tablefootnote}

\title{Voice Disorder Analysis: a Transformer-based Approach}

\name[affiliation={\clubsuit}]{Alkis}{Koudounas}
\name[affiliation={\clubsuit}]{Gabriele}{Ciravegna}
\name[affiliation={\diamondsuit\heartsuit}]{Marco}{Fantini}
\name[affiliation={\heartsuit\spadesuit}]{\\Giovanni}{Succo}
\name[affiliation={\heartsuit}]{Erika}{Crosetti}
\name[affiliation={\clubsuit}]{Tania}{Cerquitelli}
\name[affiliation={\clubsuit}]{Elena}{Baralis}

\address{
  \poli Politecnico di Torino, Turin, Italy
  \sfh San Feliciano Hospital, Rome, Italy \\
  \sgbh SCDU Otorinolaringoiatria, Head Neck Cancer Unit, Ospedale San Giovanni Bosco, Turin, Italy
  \unito Dipartimento di Oncologia, Università degli Studi di Torino, Turin, Italy
  }
\email{alkis.koudounas@polito.it, gabriele.ciravegna@polito.it}

\keywords{pathological voice disorder, transformer model, synthetic data, mixture of experts, data augmentation}

\begin{document}

\maketitle

\begin{abstract}
Voice disorders are pathologies significantly affecting patient quality of life. However, non-invasive automated diagnosis of these pathologies is still under-explored, due to both a shortage of pathological voice data, and diversity of the recording types used for the diagnosis.
This paper proposes a novel solution that adopts transformers directly working on raw voice signals and addresses data shortage through synthetic data generation and data augmentation. 
Further, we consider many recording types at the same time, such as sentence reading and sustained vowel emission, by employing a Mixture of Expert ensemble to align the predictions on different data types.
The experimental results, obtained on both public and private datasets, show the effectiveness of our solution in the disorder detection and classification tasks and largely improve over existing approaches.\footnote{This work has been accepted at Interspeech 2024.} 
\end{abstract}


\section{Introduction}

Vocal disorders are important pathologies affecting a significant portion of the population and exerting a substantial impact on patient quality of life~\cite{roy2005voice, cohen2010self, bhattacharyya2014prevalence, spantideas2015voice}. These disorders may originate from various causes, including both benign and malignant conditions, and neurodegenerative disorders~\cite{brunner2023prevalence, karabayir2020gradient, vieira2020voice}. Diagnosis often relies on clinician auditory assessments of patient voices. An early diagnosis and treatment are crucial, as they greatly improve patient prognosis. 
Thus, an automated tool that can detect and classify these disorders would be of crucial importance.

Deep learning is transforming the field of medicine by enabling new ways of diagnosing and treating diseases~\cite{rajpurkar2022ai}. One of the promising applications of deep learning is non-invasive diagnostics, which can reduce the need for invasive procedures and improve the quality of life for patients~\cite{rueda2011non}. 
However, while deep learning has made notable strides in non-invasive diagnostics for skin cancer~\cite{esteva2017dermatologist}, diabetic retinopathy~\cite{gulshan2016development}, and atrial fibrillation~\cite{attia2019screening}, voice analysis remains an under-explored domain.  
The main reasons are related to 
(i) the scarcity of voice data related to pathological issues that prevents effectively employing powerful models, and (ii) the intrinsic complexity of pathological voice data and the diversity of the data used for the diagnosis.

In this paper, we show that it is possible to overcome these challenges respectively by 
(i) generating synthetic data based on Text-to-Speech (TTS) technology and designing a strong data augmentation pipeline to enrich and balance the training data, and 
(ii) employing a Mixture of Experts (MoE) ensemble, combining Transformer models~\cite{vaswani2017attention} trained on different recording types (e.g., vowel emission and sentence readings) to capture the voice nuances available across data types.

We tested our approach on two public datasets, namely SVD~\cite{woldert2007saarbrueken} and AVFAD~\cite{jesus2017advanced}, and on an internal Italian Pathological Voice (IPV) dataset. 
The experimental validation shows that our solution significantly improves both the AUC in voice disorder detection and the F1 score in pathology classification with respect to existing models. 


\newpar{Related Works.}
Automatic voice disorder analysis has been the subject of several studies in the literature. On the one side, a few works employed multi-layer perceptron over extracted features, including acoustic features and Mel-frequency cepstral coefficients (MFCC)~\cite{salhi2008voice, arias2018byovoz}. 
On the other side, Convolutional Neural Networks (CNN) have been applied to 2D representations of the audio signal, such as the Mel spectrograms and MFCCs cepstrograms~\cite{peng2023voice, xie2023voice}. Hybrid architectures combining CNNs and Recurrent Neural Networks (RNNs) networks have also been proposed to improve the performance over long voice signal~\cite{lilhore2023hybrid}. 
Other studies have utilized models working directly on the audio signal, such as 1D-CNN~\cite{islam2022voice} and Transformers~\cite{ribas2023automatic}.
The latter is reported to boost the performance also in the related context of dysarthric speech~\cite{shahamiri2023dysarthric, almadhor2023e2e, ilias2023detecting}.   
In this paper, we use transformer models working on raw speech data as well. However, unlike~\cite{ribas2023automatic} that pre-trains the model on a separate dataset before fine-tuning it on the voice pathology detection task, we show that training on augmented and synthetic data and using an ensemble model allows us to achieve high performance.

\begin{figure*}[t]
  \centering
  \includegraphics[width=0.95\linewidth]{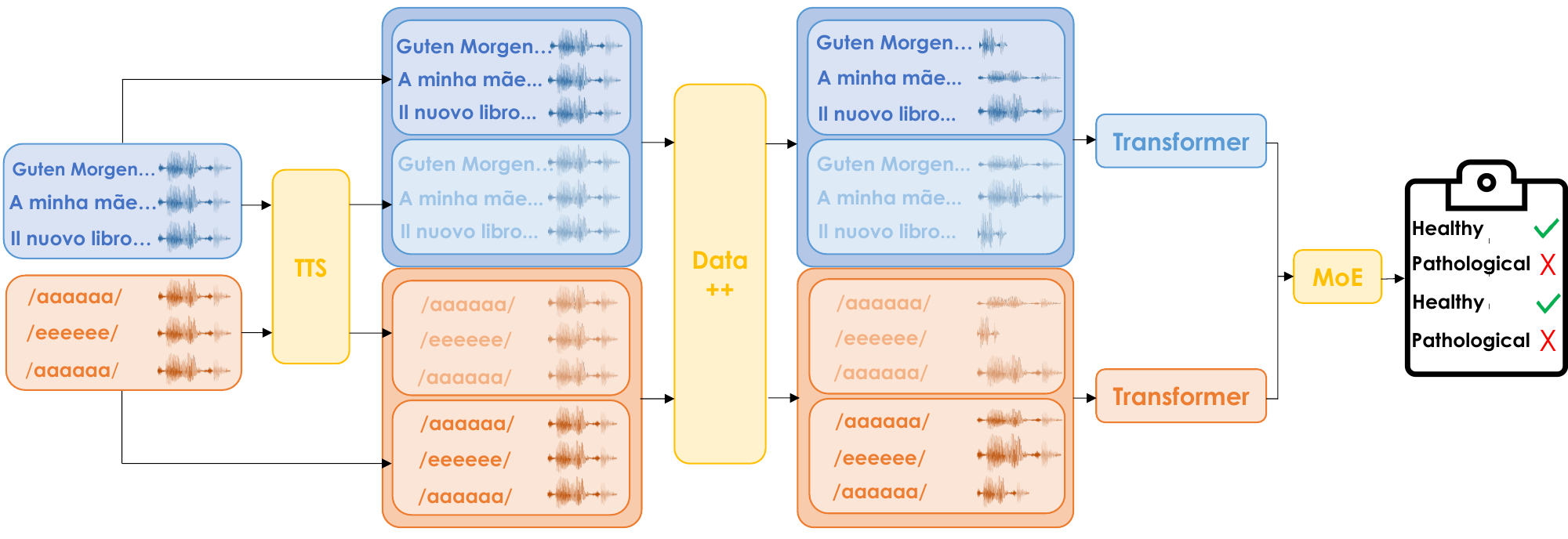}
  \caption{\textbf{Schematic diagram of our pipeline.} We synthesize (\texttt{TTS}) and augment (\texttt{data++}) the training data, separately for sentences and sustained vowels, to train two dedicated \texttt{transformer} models. We then align their predictions with a Mixture of Experts (\texttt{MoE}).}
  \label{fig:pipeline}
\end{figure*}

\section{Methodology}
We propose a pipeline to fully exploit the generalization capability of transformer models in the context of non-invasive voice disorder analysis. 
Similarly to other deep learning models, transformers require a huge quantity of data to be trained, or even fine-tuned. Moreover, each class must be sufficiently represented to avoid implicit biases.
On the contrary, publicly available datasets tend to be small and often unbalanced, exhibiting an under-representation of either pathological or healthy individuals. 
In this work, we propose to overcome this issue by generating synthetic data conditioned by the voices of actual patients, both healthy and pathological.
This approach, together with a strong standard data augmentation pipeline, aims to enhance and balance the distribution of the training datasets.

To get the most information from the available data, we also propose to consider all data types at the same time.
Indeed, the detection of a voice disorder, as well as the classification of the specific pathology causing the disorder, may benefit differently from the type of recorded sample. 
Unlike current literature focusing on a single recording type (i.e., either on sustained vowels~\cite{peng2023voice, arias2018byovoz, xie2023voice, islam2022voice} or on read sentences \cite{salhi2008voice,ribas2023automatic}), in this paper we employ a MoE ensemble, combining the predictions of transformer models analyzing different data types. In the following, we introduce the main components of the proposed pipeline, which is represented in Figure~\ref{fig:pipeline}.



\newpar{Synthetic Data Creation.}
To generate both healthy and pathological voices, we propose employing Text-to-Speech (\texttt{TTS}) and conditioning the generation process to the specific class. The TTS generation process is conditioned by incorporating the learned embeddings derived from the vocal characteristics of the real data within each class.
We employ a state-of-the-art multilingual TTS model~\cite{casanova2024xtts} that allows the automatic generation of voices for diverse datasets in different languages. 
Synthesizing voices for pathological conditions through TTS may introduce potential pitfalls, as the synthesized voice may not accurately capture the nuanced characteristics of the specific pathologies experienced by the patient. To avoid this issue, we checked the generalization capability of the synthesized voices by training a model entirely on synthetic data and testing it on the original data (see Section~\ref{sec:results}).  
To further enrich our data, we also apply a strong data augmentation pipeline (\texttt{data++}) with pitch shifting, time stretching, and noise addition. We better describe it in Section~\ref{sec:exp}. 

\newpar{Mixture of Experts.}
To exploit all available data and acknowledging the distinct characteristics of each data type in representing one's pathologies, we propose to train separate models for each input type.
We then introduce a ``shallow'' Mixture of Expert (\texttt{MoE}) framework to align the predictions of all models. 
Specifically, we select the predictions from the model providing higher confidence. 
Model confidence is estimated by the entropy of predicted probabilities: lower entropy implies higher certainty, indicating a confident prediction for the given sample.
By leveraging this integrated approach, we extract a comprehensive understanding of the pathology representation within the voice data. The experimental results show that this approach works better with respect to considering all data types within a single model.

We further consider a different pre-training for each model. 
Specifically, we propose to fine-tune on sustained vowels a model pre-trained on the Audioset dataset~\cite{audioset}, and on sentence reading a model pre-trained on the classic LibriSpeech dataset~\cite{librispeech}. 
The rationale behind this choice is based on the distinct strengths of each dataset. 
LibriSpeech is well suited for capturing the nuances of sentence readings while, following the intuition of~\cite{peng2023voice}, a model pre-trained on Audioset is expected to better represent vowels, as it comprises speech and vocalization samples, such as laughter, screaming, humming, etc. 
In the following, we will refer to this ensemble as \texttt{MoE$^*$}.

\section{Experimental Setup}
\label{sec:exp}
This section details the datasets, models, and training procedures used for the experiments.


\subsection{Datasets}
We consider two publicly-available datasets, the German \textsc{SVD}~\cite{woldert2007saarbrueken} and the Portuguese \textsc{AVFAD}~\cite{jesus2017advanced}. We also assess the performance of our approach on an internal Italian dataset (\textsc{IPV}). 
We refine public datasets to focus on similar recordings, i.e., only readings of given sentences and emissions of sustained vowels, both in a typical setting (i.e., with normal pitch).
Table~\ref{tab:datasets} summarizes the characteristics of each dataset. 
Notice that the same patients may be involved in several recordings. 

\newpar{SVD.} The Saarbruecken Voice Database (\textsc{SVD})\footnote{\href{https://stimmdb.coli.uni-saarland.de/}{https://stimmdb.coli.uni-saarland.de/}}
~\cite{woldert2007saarbrueken} includes voice recordings and electroglottography (EGG) data, with 13 files per recording session, incorporating vowels /a, i, u/ with pitch variations (normal, high, low, rising-falling) and a sentence reading task. For consistency, we only considered the sentence reading and the normal pitch vowels.
The dataset does not aggregate pathologies into macro-classes. Thus, for the disorder classification task, we considered the 6-most frequent classes.

\newpar{AVFAD.} The Advanced Voice Function Assessment Databases (\textsc{AVFAD})\footnote{\href{https://acsa.web.ua.pt/AVFAD.html}{https://acsa.web.ua.pt/AVFAD.html}}~\cite{jesus2017advanced} collect audio recordings capturing participants performing various vocal tasks, including sustaining vowels /a, e, o/, reading six sentences, reading a phonetically balanced text, and engaging in spontaneous speech, everything repeated three times.
For consistency, we concatenate the six sentences together, having three repetitions for each obtained audio. We also concatenate the sustained vowels by repetition, thus obtaining 3 audios, one for each vowel.

\newpar{IPV.} 
The Italian Pathological Voice (IPV) is a novel dataset we 
use for further testing the proposed method. The study recruited participants from several private phoniatric and speech-therapy practices and hospitals in Italy. Participants included both euphonic individuals seeking otolaryngological evaluations and participants with various degrees of dysphonia. Dysphonic participants exhibited organic and functional voice disorders of varying severity. 
Each participant underwent videolaryngostroboscopic examinations, perceptual voice evaluations, and acoustic voice analysis conducted by experienced physicians. Data collection involved two tasks: sustained production of the vowel /a/, and a reading of five phonetically balanced sentences derived from the Italian adaptation of the CAPE-V~\cite{kempster2009consensus}.
Voice samples were recorded under standardized conditions, keeping a consistent mouth-to-microphone distance of 30~cm and ensuring a quiet environment with a signal-to-noise ratio exceeding 30~dB. 
The study adhered to the principles of the Declaration of Helsinki, with all participants providing informed consent. Data analysis was conducted retrospectively and anonymously on the recorded voice samples. Further information regarding the dataset is available in the project repository\footnote{\texttt{\url{https://github.com/koudounasalkis/AI4Voice}}}.

\begin{table}[t]
  \caption{Characteristics of employed datasets (\texttt{Ds}), including language (\texttt{L}), number of healthy (\texttt{\#H}) and pathological (\texttt{\#P}) individuals, number of sentence readings (\texttt{\#S}) and sustained vowel recordings (\texttt{\#V}), number of pathological classes (\texttt{\#C}), macro classes (\texttt{\#MC}) and average audio duration (\texttt{T(s)}).}
  \label{tab:datasets}
  \centering
  \resizebox{\columnwidth}{!}{
  \begin{tabular}{lllllllll}
    \toprule
    \textbf{Ds} & \textbf{L} & \textbf{\#H} & \textbf{\#P} & \textbf{\#S} & \textbf{\# V} & \textbf{\#C} & \textbf{\#MC} & \textbf{T (s)} \\
    \midrule
        SVD~\cite{woldert2007saarbrueken}  & DE & 687 & 1356  & 2043 & 6129 & 71 & (6)\tablefootnote{As macro-classes are not given in SVD, we considered the six most frequent classes.} & 1.73  \\
        AVFAD~\cite{jesus2017advanced}     & PT & 346 & 363   & 1989 & 1989 & 25 & 8 & 15.86 \\
        IPV                                & IT & 173 & 340   & 513  & 513  & 15 & 6 & 12.89 \\
    \bottomrule
  \end{tabular}
  }
\end{table}

\subsection{Models and training procedure}

\newpar{Compared Models.} 
We reproduced the 1D and 2D CNN architectures presented in~\cite{islam2022voice, peng2023voice}, respectively.\footnote{For a fair comparison, we implemented and trained these models.} For the CNN-2D, we experimented with various MFCCs. We show the results for MFCC=40, which yielded the best outcomes overall. 
We also report the performance of vanilla transformers. 
Following~\cite{ribas2023automatic}, the suite of transformers evaluated in this work includes wav2vec 2.0~\cite{w2v2}, WavLM~\cite{wavlm}, and HuBERT~\cite{hubert}, in their base sizes. 
For wav2vec 2.0 and HuBERT, we further employ the version pre-trained on Audioset introduced in~\cite{ARCH}. 

\newpar{Training procedure.} 
We performed a 10-fold cross-validation (CV) for all the considered datasets. 
A robust data augmentation pipeline was employed to randomly replicate various noisy environments and introduce alterations in pitch (higher and lower) and time (stretched and compressed), either individually or in combination. We applied a more frequent and intense data augmentation for sentences and a slightly weaker approach for vowels, as overly aggressive augmentation for this audio type reduces the quality of the recording.
Detailed information on the hyperparameter setup can be found in the project repository.


\section{Results and Discussion}
\label{sec:results}
We tested the effectiveness of the proposed pipeline for the voice disorder detection and classification tasks against the compared methods. 
To evaluate the contribution of each part of the proposed pipeline, we also report an ablation study over each considered dataset. Finally, to assess the quality of the synthetic data, we show the results of the models when trained on synthetic data only and evaluated on real data.
All results are reported in terms of average Accuracy, F1 Macro, and AUC (for disorder detection only).  

\newpar{Voice disorder detection.}
It is a binary task, that aims at separating pathological and healthy patients.
In Table~\ref{tab:results} we report the performance on the voice disorder detection task of the proposed pipeline (\texttt{Ours}) against the compared models. The advantage of employing the proposed pipeline is significant, with +$.20$-$.36$ points in terms of AUC with respect to a 1D-CNN, +$.10$-$.26$ with respect to a 2D-CNN and +$.05$-$.13$ with respect to a plain transformer model (\texttt{HuBERT}). 



\begin{table}[ht]
    \caption{\textbf{Voice Disorder Detection}. Mean\scriptsize{ ±std} \normalsize of 10-fold CV for all datasets. Best results are in bold.} 
    \label{tab:results}
    \centering
    \begin{tabular}{clccc}
    \toprule 
    \multicolumn{1}{c}{\textbf{Ds}} & \multicolumn{1}{c}{\textbf{Approach}} & \multicolumn{1}{c}{\textbf{Accuracy}} & \multicolumn{1}{c}{\textbf{AUC}} & \multicolumn{1}{c}{\textbf{F1 Macro}} \\ 
    \midrule

    \multirow{4}{*}{\rotatebox{90}{~\parbox{0.7cm}{SVD}}}   
        & \texttt{CNN-1D}  & .746\scriptsize ±.041 & .705\scriptsize ±.041 & .722\scriptsize ±.041 \\
        & \texttt{CNN-2D}  & .799\scriptsize ±.025 & .734\scriptsize ±.025 & .747\scriptsize ±.024 \\
        & \texttt{HuBERT}  & .862\scriptsize ±.040 & .844\scriptsize ±.041 & .842\scriptsize ±.038 \\
        & \texttt{Ours}    & \textbf{.909\scriptsize ±.006} & \textbf{.911\scriptsize ±.005} & \textbf{.907\scriptsize ±.007} \\
    \midrule

    \multirow{4}{*}{\rotatebox{90}{~\parbox{1.1cm}{AVFAD}}}   
        & \texttt{CNN-1D} & .712\scriptsize ±.028 & .719\scriptsize ±.029 & .711\scriptsize ±.029 \\
        & \texttt{CNN-2D} & .835\scriptsize ±.019 & .834\scriptsize ±.021 & .834\scriptsize ±.021 \\
        & \texttt{HuBERT} & .872\scriptsize ±.015 & .877\scriptsize ±.015 & .871\scriptsize ±.014 \\
        & \texttt{Ours}   & \textbf{.927\scriptsize ±.004} & \textbf{.931\scriptsize ±.004} & \textbf{.926\scriptsize ±.004} \\
    \midrule

    \multirow{4}{*}{\rotatebox{90}{~\parbox{0.6cm}{IPV}}}   
        & \texttt{CNN-1D} & .673\scriptsize ±.025 & .616\scriptsize ±.024 & .637\scriptsize ±.025 \\
        & \texttt{CNN-2D} & .788\scriptsize ±.021 & .721\scriptsize ±.021 & .737\scriptsize ±.021 \\
        & \texttt{HuBERT} & .875\scriptsize ±.024 & .847\scriptsize ±.026 & .870\scriptsize ±.026 \\
        & \texttt{Ours}   & \textbf{.981\scriptsize ±.005} & \textbf{.983\scriptsize ±.006} & \textbf{.978\scriptsize ±.005} \\
    \bottomrule
    
    \end{tabular}
    
\end{table}

\newpar{Voice disorder classification.}
This multilabel classification task aims at distinguishing among different pathologies causing the disorder (e.g., polyps, nodules, cysts). 
In Table~\ref{tab:results_category} we report the results for voice disorder classification. 
The performance gap becomes very significant, particularly when considering the F1 macro, which takes into account class imbalance. 
Our method reports an increase for the F1 score of +$.48$-$.57$ with respect to the CNN-1D, +$.38$-$.52$ with respect to the CNN-2D, +$.11$-$.15$ with respect to a plain transformer.  

We also propose leveraging an initial pre-training on the voice disorder detection task to further enhance the performance on the classification task. We ensured that, in the first training, the model did not encounter the speakers included in the test set for the classification task. We call this approach \texttt{ours$^*$}.
As expected, the pre-training boosts the final performance, with a further increase of +$.01$-$.05$ over the version without pre-training.

These results show that the employment of synthetic data and data augmentation, together with an ensemble model to address different data types, fully exploits the generalization capability of transformer models. In the following, we will analyze the contribution of the different pipeline components through an ablation study. 
\begin{table}[ht]
    \caption{\textbf{Voice Disorder Classification.} Mean\scriptsize{ ±std} \normalsize of 10-fold CV for \textsc{AVFAD} and \textsc{IPV} datasets. Best results are in bold.} 
    \label{tab:results_category}
    \centering
    \begin{tabular}{clcc}
    \toprule 
    \multicolumn{1}{c}{\textbf{Ds}} & \multicolumn{1}{c}{\textbf{Model}} & \multicolumn{1}{c}{\textbf{Accuracy}} & \multicolumn{1}{c}{\textbf{F1 Macro}} \\ 
    \midrule

    \multirow{5}{*}{\rotatebox{90}{~\parbox{0.7cm}{SVD}}}   
        & \texttt{CNN-1D}  & .437\scriptsize±.025  & .280\scriptsize±.024 \\
        & \texttt{CNN-2D}  & .539\scriptsize±.021  & .348\scriptsize±.023 \\
        & \texttt{HuBERT}  & .771\scriptsize±.022  & .712\scriptsize±.020 \\
        & \texttt{Ours}    & .874\scriptsize±.017  & .859\scriptsize±.014 \\
        & \texttt{Ours$^*$}& \textbf{.888\scriptsize±.015} & \textbf{.868\scriptsize±.016} \\
    \midrule
    
    \multirow{5}{*}{\rotatebox{90}{~\parbox{1.1cm}{AVFAD}}}   
        & \texttt{CNN-1D}  & .401\scriptsize±.027  & .167\scriptsize±.028 \\
        & \texttt{CNN-2D}  & .509\scriptsize±.025  & .266\scriptsize±.024 \\
        & \texttt{HuBERT}  & .693\scriptsize±.028  & .538\scriptsize±.026 \\
        & \texttt{Ours}    & .782\scriptsize±.024  & .648\scriptsize±.023 \\
        & \texttt{Ours$^*$}& \textbf{.808\scriptsize±.021} & \textbf{.703\scriptsize±.020} \\
    \midrule
    
    \multirow{5}{*}{\rotatebox{90}{~\parbox{0.6cm}{IPV}}}   
        & \texttt{CNN-1D}  & .419\scriptsize±.022  & .278\scriptsize±.024 \\
        & \texttt{CNN-2D}  & .521\scriptsize±.019  & .335\scriptsize±.021 \\
        & \texttt{HuBERT}  & .764\scriptsize±.025  & .710\scriptsize±.023 \\
        & \texttt{Ours}    & .867\scriptsize±.010  & .854\scriptsize±.007 \\
        & \texttt{Ours$^*$}& \textbf{.883\scriptsize±.007} & \textbf{.871\scriptsize±.006} \\
    \bottomrule
    
    \end{tabular}
    
\end{table}


\newpar{Ablation study.}
We report in Table~\ref{tab:ablation_study} the AUC performance of our model when adding the different elements of the proposed pipeline. More precisely, we tested the improvement given by adding to a transformer base model trained on the sentence reading data type\footnote{Please note that the base model trained on sustained vowels achieves worse performance w.r.t. its counterpart on sentences, thus we only report it in the project repository.} (\texttt{base}), the data augmentation (\texttt{data++}), and the synthetic data (\texttt{TTS}). We then tested the employment of both sentence reading and sustained vowels with an ensemble model (\texttt{MoE}). 
By further differentiating the pre-training of each model on ad-hoc datasets (LibriSpeech for sentence reading, Audioset for sustained vowels), we obtain our model (\texttt{MoE$^*$}).\footnote{A pre-trained version of WavLM on the AudioSet dataset is not publicly available. Thus the \texttt{MoE$^*$} result is not reported in this case. } 

We can observe from Table~\ref{tab:ablation_study} that the data augmentation and the synthetic data creation yield significant improvements, albeit with varying degrees across datasets and models, ranging from +$0.01$ on SVD to +$0.10$ on IPV. Conversely, employing an ensemble model (with ad-hoc pre-training) consistently enhances the performance by approximately +$0.04$ across all datasets and models.
Employing all data types at the same time for a single model (\texttt{ALL}) results instead in a bad strategy, decreasing the performance of the model below the baseline. These results confirm our intuition that different data types provide complementary information that can improve prediction accuracy. At the same time, their diversity prevents considering them together, and requires separate model training for each data type.  

\begin{table}[ht]
    \caption{\textbf{Ablation study} on voice disorder detection to quantify the contribution of each term in terms of AUC. Best results for each model in bold, best results overall in \colorbox[HTML]{C2F0C2}{light-green}.
    } 
    \label{tab:ablation_study}
    \centering
    \begin{tabular}{clccc}
    \toprule
    
    \textbf{Ds} & \textbf{Approach} & \textbf{HuBERT} & \textbf{wav2vec 2.0} & \textbf{WavLM} \\
    \midrule
    
    \multirow{6}{*}{\rotatebox{90}{~\parbox{0.7cm}{SVD}}} 
       & \texttt{base}                & .844\scriptsize±.041 & .842\scriptsize±.038 & .842\scriptsize±.035 \\
       & \texttt{+ data++}            & .851\scriptsize±.032 & .849\scriptsize±.036 & .849\scriptsize±.032 \\
       & \texttt{+ TTS}               & .871\scriptsize±.051 & .855\scriptsize±.032 & .859\scriptsize±.039 \\
       & \texttt{+ MoE}               & .903\scriptsize±.012 & .888\scriptsize±.013 & \textbf{.884\scriptsize±.019} \\
       & \texttt{+ MoE}$^*$           & \colorbox[HTML]{C2F0C2}{\textbf{.911\scriptsize±.005}} & \textbf{.894\scriptsize±.011} & -           \\
       \cmidrule{2-5}
       & \texttt{ALL}                 & {.791\scriptsize±.031} & .787\scriptsize±.029           & .784\scriptsize±.038 \\
     \midrule
     
     \multirow{6}{*}{\rotatebox{90}{~\parbox{1.1cm}{AVFAD}}} 
       & \texttt{base}                & .877\scriptsize±.015 & .875\scriptsize±.018 & .872\scriptsize±.016 \\
       & \texttt{+ data++}            & .889\scriptsize±.019 & .879\scriptsize±.017 & .881\scriptsize±.014 \\
       & \texttt{+ TTS}               & .894\scriptsize±.015 & .882\scriptsize±.015 & .885\scriptsize±.019 \\
       & \texttt{+ MoE}               & .917\scriptsize±.007 & .902\scriptsize±.014 & \textbf{.908\scriptsize±.007} \\
       & \texttt{+ MoE}$^*$           & \colorbox[HTML]{C2F0C2}{\textbf{.931\scriptsize±.004}} & \textbf{.921\scriptsize±.009} & -             \\
       \cmidrule{2-5}
       & \texttt{ALL}                 & {.865\scriptsize±.012} & .861\scriptsize±.017           & .863\scriptsize±.013 \\
     \midrule
     
     \multirow{6}{*}{\rotatebox{90}{~\parbox{0.6cm}{IPV}}} 
       & \texttt{base}                & .847\scriptsize±.026 & .874\scriptsize±.021 & .832\scriptsize±.029  \\
       & \texttt{+ data++}            & .903\scriptsize±.016 & .894\scriptsize±.019 & .845\scriptsize±.022  \\
       & \texttt{+ TTS}               & .948\scriptsize±.021 & .933\scriptsize±.021 & .929\scriptsize±.020  \\
       & \texttt{+ MoE}               & .977\scriptsize±.013 & .943\scriptsize±.015 & \textbf{.930\scriptsize±.012}  \\
       & \texttt{+ MoE}$^*$           & \colorbox[HTML]{C2F0C2}{\textbf{.983\scriptsize±.006}} & \textbf{.970\scriptsize±.005}           & -              \\
       \cmidrule{2-5}
       & \texttt{ALL}                 & {.831\scriptsize±.015} & .818\scriptsize±.018           & .787\scriptsize±.016              \\
    \bottomrule
    
    \end{tabular}
\end{table}

\newpar{Generalization performance of synthetic data.}
As previously observed, 
while altering and creating more variety in the training distribution by means of synthetic data, we still need to preserve the characteristics of the original data. 
We assessed the quality of the synthetic data by training the models only on the synthetic data and checking their performance on the real data. 
In this case, our model, while not incorporating additional synthetic data, integrates the HuBERT model with data augmentation and the MoE ensemble technique.
In Table~\ref{tab:results_tts_only} we present the results 
on the IPV dataset. 
The drop in accuracy, compared to using solely real data, ranges from -$.07$ to -$.14$ in detection and -$.05$ to -$.12$ in classification for both the CNN and transformer approaches. This small decline underscores the quality of synthetic data in emulating real data characteristics, as model performance remain significantly above random chance and close to the original performance. 
\begin{table}[ht]
    \caption{\textbf{Synthetic Data Only, IPV dataset.} Mean\scriptsize{ ±std} \normalsize of 10-fold CV. Best results in bold. Difference with real data in brackets.} 
    \label{tab:results_tts_only}
    \centering
    \begin{tabular}{clcc}
    \toprule 
    \multicolumn{1}{c}{\textbf{Task}} & \multicolumn{1}{c}{\textbf{Model}} & \multicolumn{1}{c}{\textbf{Accuracy}} & \multicolumn{1}{c}{\textbf{F1 Macro}} \\ 
    \midrule

    \multirow{4}{*}{\rotatebox{90}{~\parbox{1.35cm}{Detection}}}   
        & \texttt{CNN-1D}   & .573{\scriptsize±.029} \ \ (-.100) & .564{\scriptsize±.030} \ \ (-.073)\\
        & \texttt{CNN-2D}   & .659{\scriptsize±.031} \ \ (-.129) & .592{\scriptsize±.033} \ \ (-.145)\\
        & \texttt{HuBERT}   & .794{\scriptsize±.028} \ \ (-.081) & .782{\scriptsize±.025} \ \ (-.088)\\
        & \texttt{Ours}     & \textbf{.868\scriptsize±.012}\ \  (-.113)& \textbf{.854\scriptsize±.011} \ \  (-.124) \\
    \midrule

    \multirow{5}{*}{\rotatebox{90}{~\parbox{1.75cm}{ Classification}}}   
        & \texttt{CNN-1D}   & .308{\scriptsize±.034} \ \ (-.111)& .221{\scriptsize±.036} \ \ (-.057)\\
        & \texttt{CNN-2D}   & .387{\scriptsize±.033} \ \ (-.234)& .284{\scriptsize±.032} \ \ (-.051)\\
        & \texttt{HuBERT}   & .651{\scriptsize±.027} \ \ (-.113)& .622{\scriptsize±.026} \ \ (-.088)\\
        & \texttt{Ours}     & .749{\scriptsize±.025} \ \ (-.118)& .743{\scriptsize±.026} \ \ (-.111)\\ 
        & \texttt{Ours$^*$} & \textbf{.755\scriptsize±.021} \ \ (-.128) & \textbf{.751\scriptsize±.022} \ \ (.-120)\\ 
    \bottomrule
    
    \end{tabular}
    
\end{table}

\section{Conclusion}
We proposed a novel approach to accurately detect and classify voice disorders.
Its experimental evaluation shows that using a strong data synthesis and augmentation strategy, and employing an ensemble of transformer models 
allows achieving significant performance improvements on both tasks.


\newpar{Limitations and future work.}
The main limitations of our approach are related to (i)~the model size and (ii)~the real-world generalization of the model. For (i), while the MoE boosts the performance over a model trained on a single data type, it doubles the model size. 
A possible approach to weight sharing could entail employing a single transformer working on both data types at the same time. 
For (ii), the models have been trained on data recorded under expert guidance. As future work, we would like to test and extend our model by considering real-world scenarios. To this aim, we are embedding our model in a web-based application that will be deployed in a number of selected private and public otolaryngology clinics.

\section{Acknowledgements}
This work is partially supported by the FAIR - Future Artificial Intelligence Research and received funding from the European Union Next-GenerationEU (PIANO NAZIONALE DI RIPRESA E RESILIENZA (PNRR) – MISSIONE 4 COMPONENTE 2, INVESTIMENTO 1.3 – D.D. 1555 11/10/2022, PE00000013). 
This manuscript reflects only the authors' views and opinions, neither the European Union nor the European Commission can be considered responsible for them. 

\bibliographystyle{IEEEtran}
\bibliography{template}

\end{document}